\documentclass[multphys,vecphys]{svmult}

\usepackage{makeidx}         
\usepackage{graphicx}        
\usepackage{multicol}        
\usepackage[bottom]{footmisc}

\makeindex             

\begin{document}

\title*{Hot Massive Stars: The Impact of HST}

\titlerunning{Hot Massive Stars}

\author{Paul A. Crowther}

\authorrunning{Crowther}

\institute{Dept of Physics \& Astronomy, University of Sheffield,
Hounsfield Road, Sheffield, S3 7RH, United Kingdom
\texttt{Paul.Crowther@sheffield.ac.uk}}
%
%
\maketitle

\begin{abstract}
We review the contribution of Hubble Space Telescope to the study of hot, 
luminous stars. Optical and IR imaging have permitted spatially resolved 
observations of young, massive clusters within Local Group galaxies, such 
as R136, NGC~3603 and Arches, revealing unprecedented concentrations of 
very massive O stars. UV spectroscopy of field OB stars in the 
Magellanic Clouds have provided suitable templates for interpretation of 
metal-poor star-forming  galaxies at high-redshift. Spectacular imaging 
provides the detailed structure of ejecta nebulae from individual stars, 
including the Homunculus associated with $\eta$ Carinae and M1--67, 
associated with a Wolf-Rayet star. HST has permitted individual massive
stars to be spatially resolved in giant HII regions located beyond the 
Local Group, such as NGC~604, plus individual clusters, dominated 
by the light of massive stars within starburst galaxies at larger 
distances, such as NGC~3125.  UV spectroscopy of young, massive 
clusters in the extremely metal-poor HII galaxy I\,Zw~18 include 
signatures of large numbers of Wolf-Rayet stars.
 \end{abstract}

\section{Introduction}\label{s1}

Massive stars distinguish themselves from their lower mass siblings by 
their exceptionally high main-sequence luminosities, such that their 
lifetimes are measured in Myr rather than Gyr. Individual stars may be 
readily studied in detail in external galaxies - it was not be accident 
that the first post-servicing mission WFPC2 image was obtained of the 
Wolf-Rayet star Melnick~34 in the 30 Doradus star-forming region of the 
LMC (News Release: STScI-1994-05). Here, selected HST results are 
discussed, together with their impact 
upon European astronomy.

\section{Stellar Winds - Metallicity dependent winds}\label{s2}

The first balloon and satellite missions provided the means of studying
winds from hot stars, via ultraviolet P~Cygni profiles, for which IUE
provided a comprehensive sample of Milky Way OB stars (e.g. Howarth \& 
Prinja 1989\nocite{hp89}). Comparable quality spectroscopy of OB stars in 
the 
metal-poor Magellanic Clouds required the superior throughput and spatial 
resolution of HST (Walborn et al. 1995\nocite{wal95}, 2000\nocite{wal00}, 
Evans et al. 2004\nocite{eva04}).

Winds of early-type stars are predicted to be driven by radiation pressure 
through (CNO, Fe-peak) metal-lines (e.g. Vink et al. 2001\nocite{vin01}), 
from which 
metal-poor stars are expected to possess lower density, slower winds. 
Indeed, wind velocities of early O stars in the SMC were established to be 
slower than those of comparable Galactic stars (e.g. Prinja \& Crowther 
1998\nocite{pc98}), from CIV 1550 observations, although it has taken 
large 
ground-based surveys, such as the 
VLT/FLAMES survey of massive stars in Milky Way, LMC and SMC clusters to 
quantify the metallicity-dependence of mass-loss rates, revealing dM/dt 
$\propto Z^{0.78 \pm 0.17}$ (Mokiem et al. 2007\nocite{mok07}), in good 
agreement with 
theory $Z^{0.69 \pm 0.10}$ according to Vink et al. (2001\nocite{vin01}).

\section{Ejecta nebulae - Signatures of mass ejections}\label{s3}

Amongst the many images obtained with HST, one of the most breathtaking 
has been the WFPC2 images of Morse et al. (1998)\nocite{mor98} of the 
Homunculus 
reflection nebula, produced by the `eruption' of the prototype 
Luminous Blue Variable (LBV) $\eta$ Carinae during the 
mid-19th Century, and now illuminated from within. The expanding emission 
lobes extend  8.5$''$ (0.1 pc)  from the central star, now known to be a 
5.5~yr period binary system, for which WFPC2 achieved a  spatial 
resolution of $\sim$115~AU, comparable with the size of our Solar System.
A physical mechanism for the eruption, in which in excess of 10$M_{\odot}$ 
were ejected over 20 years, remains unclear.  STIS  
long-slit spectroscopy  of the central star,  reveals exceptional 
properties, with current mass-loss rates of 10$^{-3} M_{\odot}$ yr$^{-1}$
for an adopted (infrared) luminosity of $5\times 10^{6} L_{\odot}$ 
(Hillier et al. 2001).\nocite{hil01}

\begin{figure}
\includegraphics[width=1cm]{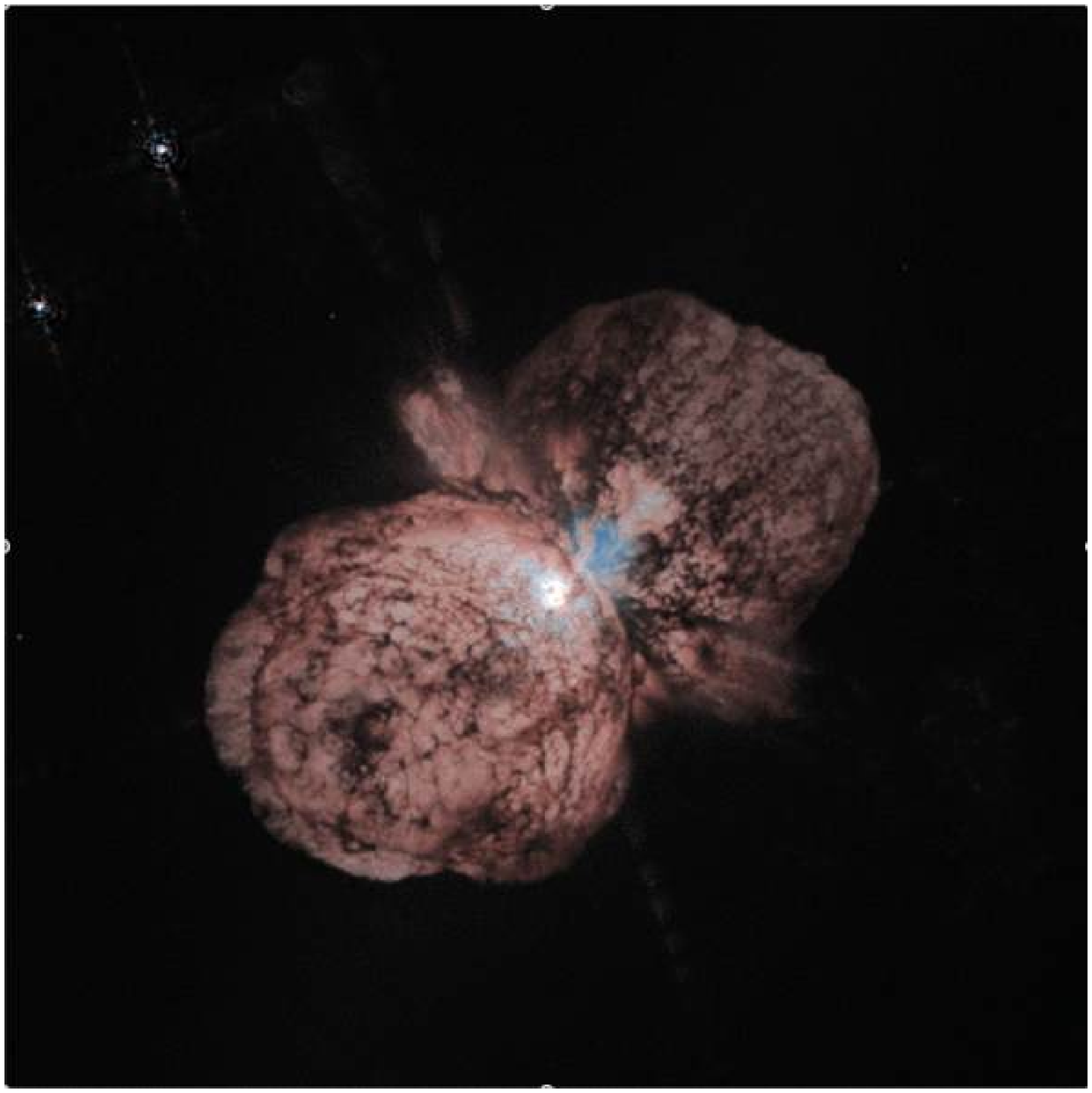}
\hspace{\fill}
\includegraphics[width=10cm]{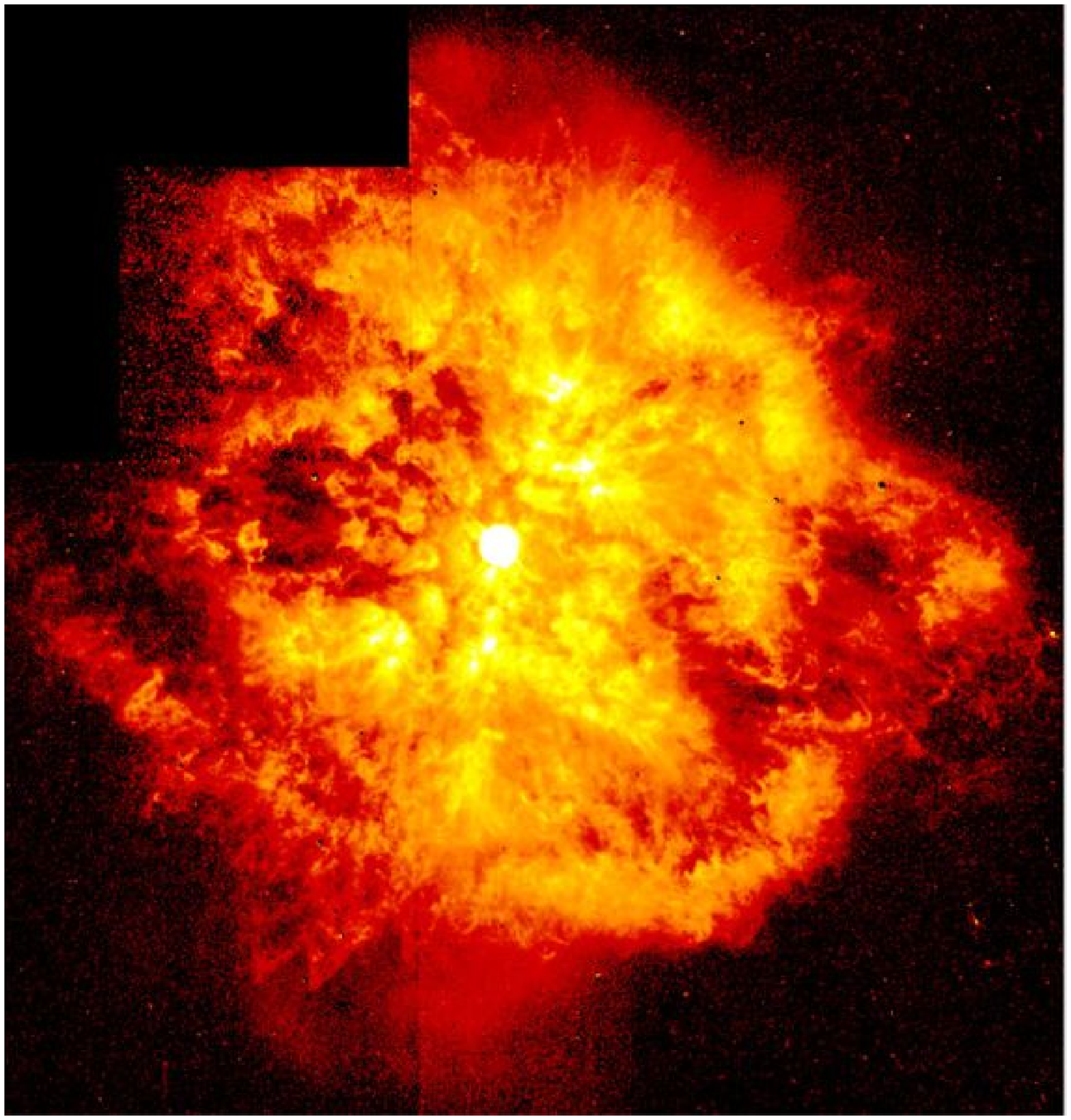}
\caption{WFPC2 imaging, to scale, of the Homunculus nebula associated with
$\eta$ Carinae (left, 0.3$\times$0.3~pc, STScI-1996-23,
Morse et al.  1998)  and M1--67, associated with a Wolf-Rayet 
star (right, 3$\times$3~pc, STScI-1998-38,
see Grosdidier et al. 1998).}\nocite{gro98}\nocite{mor98}
\label{f1}
\end{figure}


WFPC2 has also provided an unprecedented view of the 
young ejecta (or ring) nebula M1--67 associated with the Galactic 
Wolf-Rayet star WR124 (Grosdidier et al. 1998).\nocite{gro98} The radial 
density 
distribution of this nebula, approximately 90$''$ ($\sim$2~pc) in diameter 
with a $r^{-0.7}$ dependence, has enabled photo-ionization modelling of 
the central WR star (Crowther et al. 1999)\nocite{cro99} and may represent 
the immediate 
environment into which cosmological Gamma-Ray Bursts (GRBs) explode.
M1--67 is compared to $\eta$ Carinae on a common physical scale in 
Figure~\ref{f1}.

\section{Massive binaries - Colliding winds}\label{s4}

High spatial resolution radio surveys of massive binaries reveal thermal
(stellar wind) and non-thermal (colliding wind) components (e.g. Williams
et al. 1997).\nocite{wil97} Positions of stars within such systems have 
been established 
with WFPC2 imaging, enabling their relative wind strengths to be 
established (Niemela et al. 1998).\nocite{nie98} 

Indeed, FGS has enabled searches 
for hitherto unknown massive binaries in regions of parameter space 
inaccessible to ground-based techniques. One such survey of 23 OB
stars in Carina revealed 5 new binaries, including an apparent
early O  dwarf
companion to HD~93129A, the prototype O2 supergiant, separated by
only 55 mas (137~AU, Nelan et al. 2004).\nocite{nel04} Binarity was later 
confirmed by 
detection of a  non-thermal component in the observed radio emission.

To date, the current heavyweight record holder is WR20a, a 3.7 day 
eclipsing binary system composed of two H-rich WN-type stars each of
$\sim 82M_{\odot}$ (Rauw et al. 2005).\nocite{rau05}

\section{Young star clusters - A plethora of hot stars}\label{s5}

Historically, R136 the central ionizing cluster of 30 Doradus 
in the LMC was 
considered
as a potential supermassive star. 30 Doradus is the brightest
giant HII region within the Local Group, responsible for the equivalent
of 1,000 equivalent O7\,V stars. Weigelt \& Baier (1985)\nocite{wb85} 
first resolved the central source R136 into multiple components using
speckle imaging. De Marchi et al. (1993)\nocite{dem93} provided 
confirmation
with FOC, while Massey \& Hunter (1998)\nocite{mh98}
undertook FOS
spectroscopy of individual sources within R136, revealing a
multitude of early O stars, indicating extreme youth (1--2 Myr)
and apparently very high individual stellar masses, perhaps up
to 120 $M_{\odot}$. The total stellar mass of R136 probably exceeds
$5\times 10^{4} M_{\odot}$ (Hunter et al. 1996\nocite{hun96}). 
NICMOS  imaging of 30 Doradus
has identified still younger, visibly obscured, young massive stars.

\begin{figure}
\includegraphics[width=3.15cm]{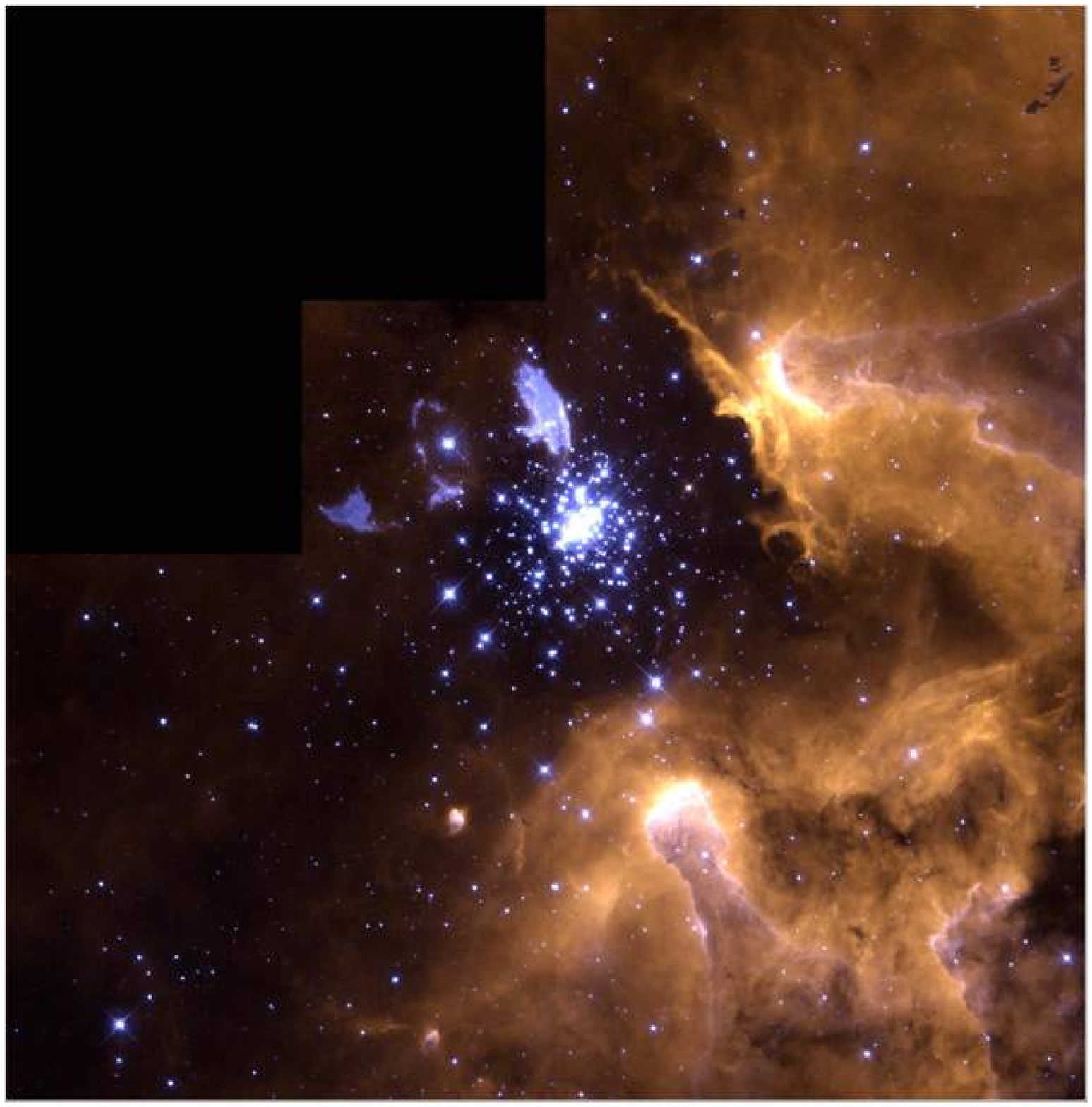}
\hspace{\fill}
\includegraphics[width=7.85cm]{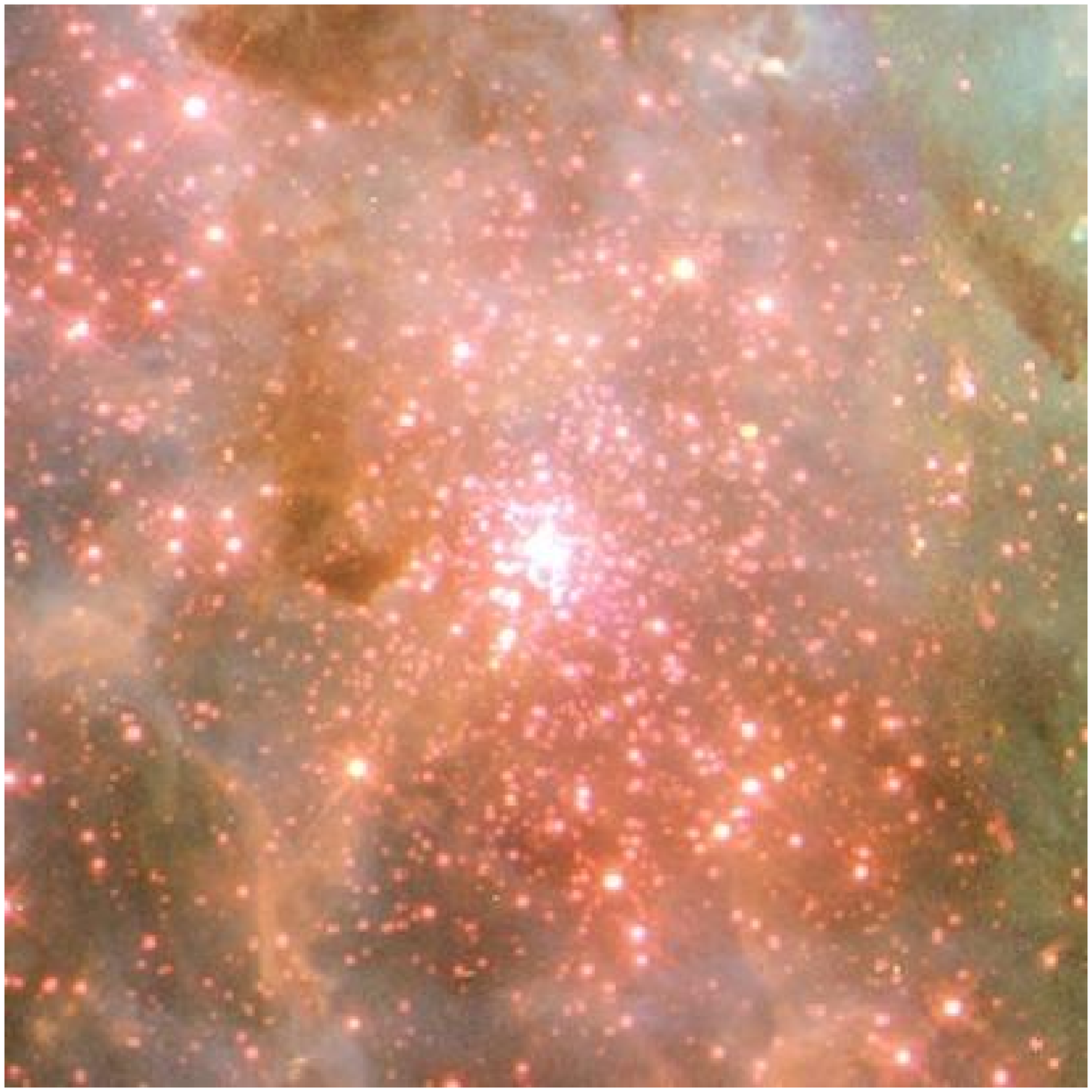}
\caption{WFPC2 imaging, to scale, of the young star clusters NGC~3603
(4$\times$4\,pc, STScI-1999-20, Brandner et al. 2000) and R136 
(10$\times$10\,pc, ESA heic0416, Hunter et al. 1996)}
\label{f2}\nocite{hun96}\nocite{bra00}
\end{figure}


HST has also spatially resolved the Milky Way cluster NGC~3603,
again revealing many early O stars (Drissen et al. 1995).\nocite{dri95} 
The 
central 
cluster is comparable to R136a within a radius of $\sim$1pc (Crowther \& 
Dessart 1998)\nocite{cd98}. Elsewhere in the Milky Way, NICMOS has 
been used to 
spatially resolve the compact ($<$0.5~pc) Arches cluster close to the 
Galactic centre. The 
Arches cluster is apparently sufficiently young that the present-day mass 
function approximates the true IMF, from which an upper stellar mass limit 
of $\sim$150 $M_{\odot}$ has been proposed (Figer 2005).\nocite{fig05}

\begin{figure}
\includegraphics[width=2.2cm]{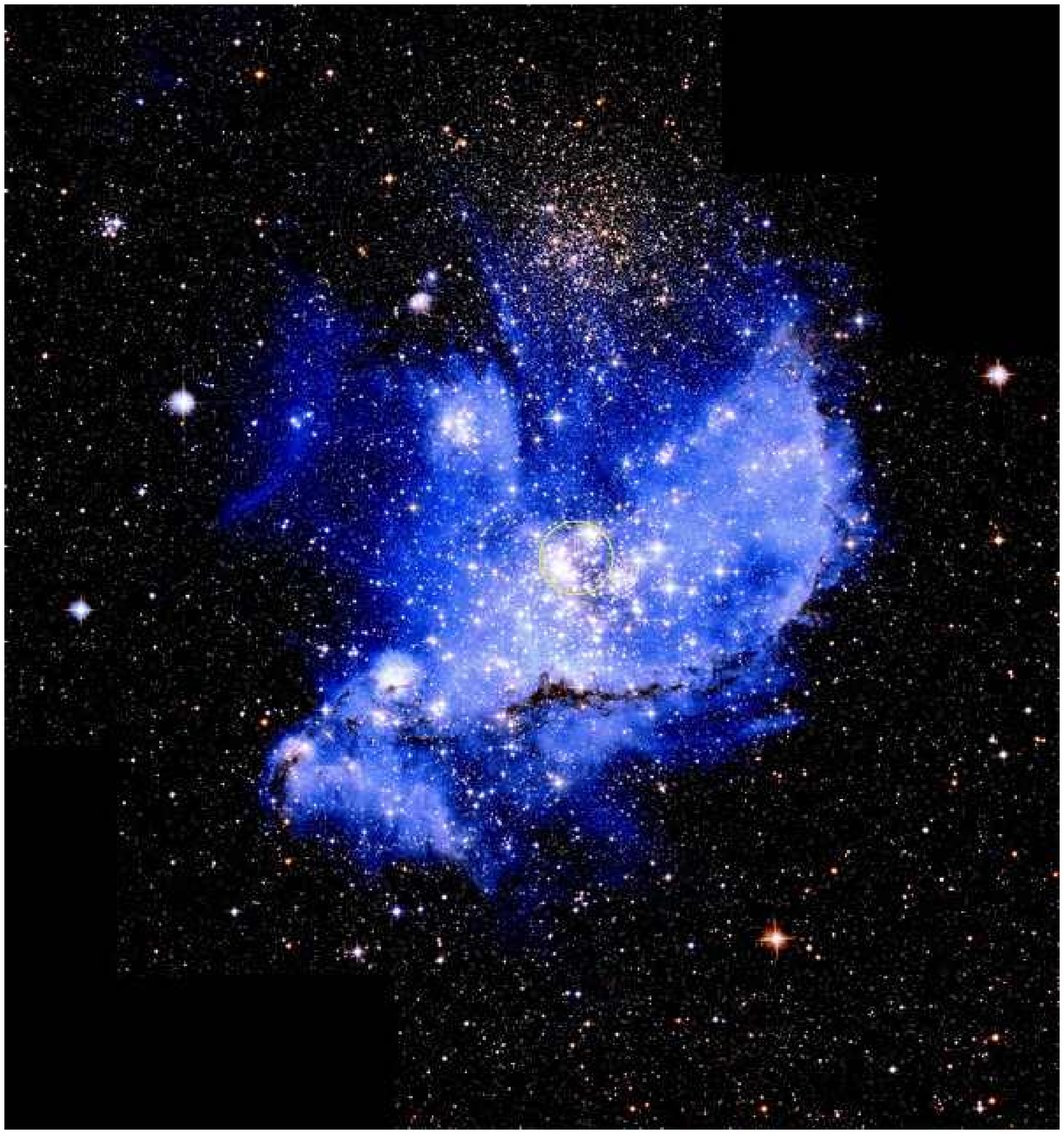}
\hspace{\fill}
\includegraphics[width=8.8cm]{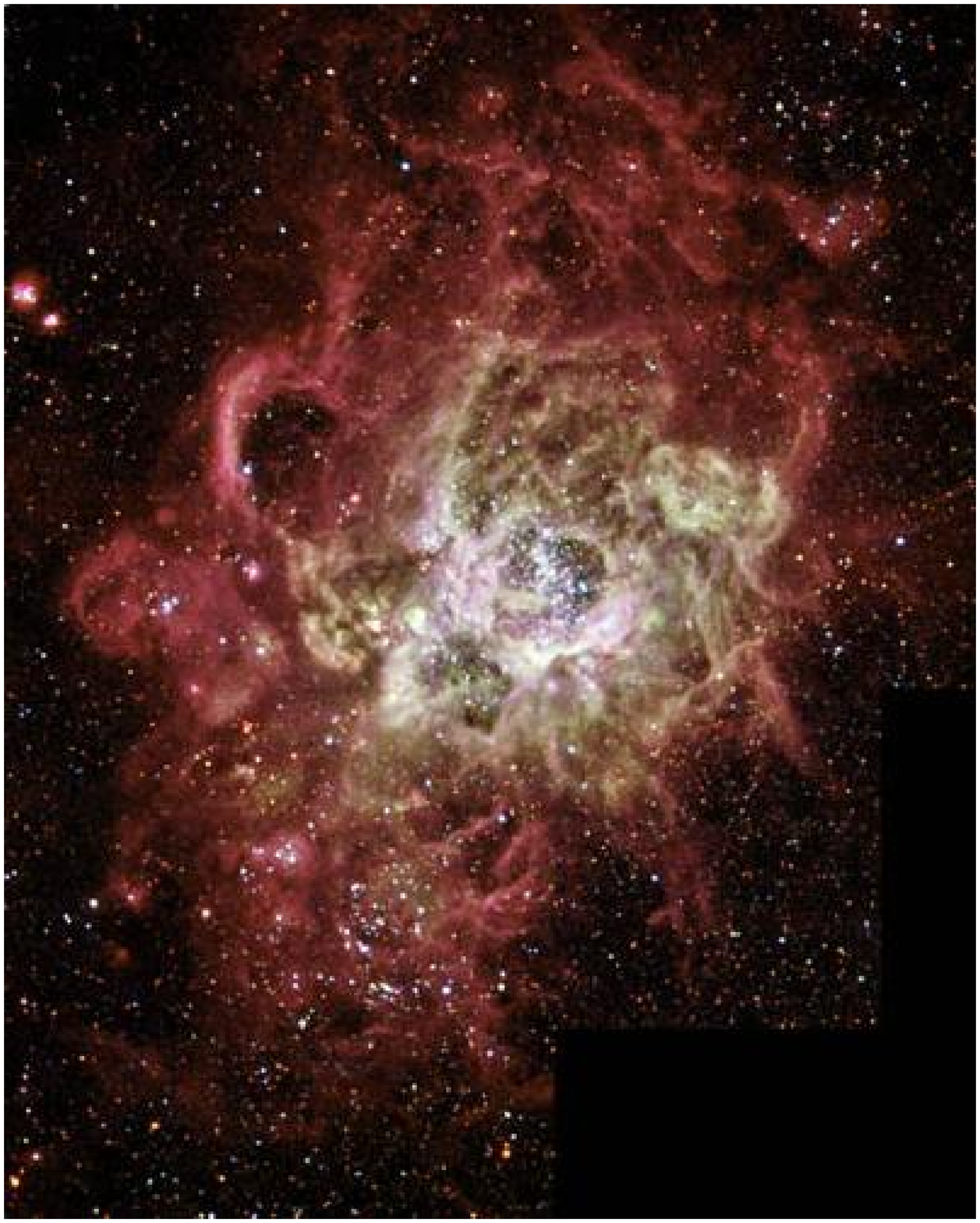}
\caption{HST imaging, to scale, of the giant HII regions
NGC~346 (SMC N11A) in the SMC 
(ACS: 90$\times$90\,pc, STScI-2005-04, 
Nota et al. 2006)  and 
NGC~604 in M~33 (WFPC2: 360$\times$450\,pc, 
STScI-1996-27: Yang et al. 1996)}
\label{f3}
\nocite{not06}\nocite{yan96}
\end{figure}

At present, the most massive young Milky Way cluster known is Westerlund~1 
(Wd~1), in which $\sim 5 \times 10^{4} M_{\odot}$ are contained within a 
radius of $\sim$1.5~pc (Clark et al. 2005).\nocite{cla05} This cluster 
hosts a full 
menagerie of rare, massive stars, including Wolf-Rayet stars, red 
supergiants, yellow hypergiants, an LBV and a B[e] supergiant.

Such high mass, compact clusters are unusual for normal spiral galaxies. 
Large, scaled-up OB associations are more typical, for which WFPC2 
imaging provided an excellent example of NGC~604 in M33 (see
Yang et al. 1996\nocite{yan96}), several hundred
pc in diameter, containing the equivalent of several hundred
O7\,V stars.  This is compared to ACS imaging of  NGC~346 in the SMC (Nota 
et al. 2006\nocite{not06}) in Fig.~\ref{f3}, somewhat smaller and
with the equivalent of 50 equivalent O7\,V stars (Kennicutt 
1984).\nocite{ken84} 
The high spatial resolution achieved with WFPC2 
permitted Drissen et al. (1993)\nocite{dri93} to identify individual 
Of and Wolf-Rayet emission line stars 
in NGC~604 to be established. A similar ground-based study of 
Wolf-Rayet stars within a bright giant HII region in the southern barred 
spiral galaxy NGC~1313 by Hadfield \& Crowther (2007)\nocite{hc07},
comparable in size and ionizing output to NGC~604
demonstrates the 
invaluable role of ACS in disentangling the stellar content. Field 
Wolf-Rayet stars at the 4~Mpc distance of NGC~1313 can be detected from 
ground-based telescopes, but HST has proved decisive in detection 
core-collapse supernova progenitors from pre-SN imaging (e.g. Smartt et 
al. 2004).\nocite{sma04}

\begin{figure}
\centering
\includegraphics[height=8cm]{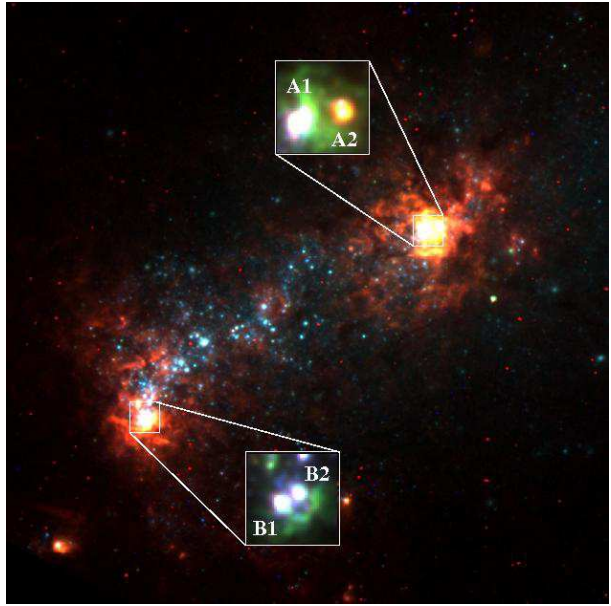}
\caption{20$\times$20$''$ ACS imaging of the blue compact dwarf
galaxy NGC~3125 (U, V, H$\alpha$,  GO\#10400, R.~Chandar)
in  which young massive ($\sim 10^{5} M_{\odot}$) clusters from knots A 
and B, $\sim$500 pc apart (Hadfield \& Crowther 2006) are 
indicated.}\nocite{hc06}
\label{f4}       
\end{figure}

\section{Starburst knots - Templates for high-$z$ galaxies}\label{s6}

\begin{figure}
\centering
\includegraphics[height=9cm,angle=90]{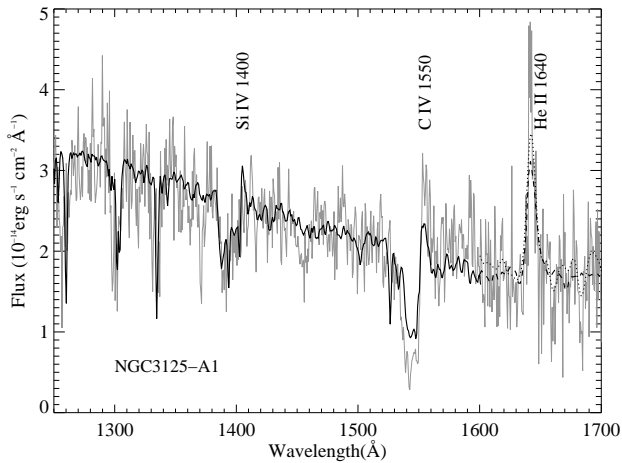}
\caption{Dereddened, slit-loss corrected STIS spectroscopy of NGC~3125--A1 
(thin solid line), 
together with a 2$\times 10^{5} M_{\odot}$ Starburst99 Magellanic Cloud 
4\,Myr instantaneous burst synthetic
spectrum (thick solid line), plus LMC WN5--6 templates at He\,{\sc ii} 
$\lambda$1640 (dashed-line) from Hadfield \& Crowther (2006).}\nocite{hc06}
\label{f5}       
\end{figure}

Local starbursts, such as NGC~3125 (11~Mpc) host young massive clusters,
significantly more massive than R136, as presented in Figure~\ref{f4}. 
Starbursts from such galaxies provide an order of magnitude higher
ionizing output than NGC~604 from within a much smaller region, e.g.
the equivalent of 2,500 equivalent O7\,V stars from knot A of NGC~3125.
STIS ultraviolet 
spectroscopy of individual clusters enables robust age estimates of such
clusters, based on comparisons between the OB P~Cygni features and 
spectral synthesis predictions (e.g. Chandar et al. 2004)\nocite{cha04}, 
as shown in Figure~\ref{f5} for NGC~3125-A1 from Hadfield \& Crowther 
(2006)\nocite{hc06}. Host galaxies typically possess Magellanic Cloud 
metallicities, such that the use of template LMC/SMC OB stars (recall 
Sect.\ref{s2}) has proved to be invaluable. 

In extreme cases, far lower metallicities are sampled, such as the HII 
galaxy I\,Zw~18, possessing 1/30 $Z_{\odot}$ based upon the recently 
reduced 
Solar oxygen abundance. Brown et al. (2002)\nocite{bro02} undertook 
STIS ultraviolet 
spectroscopy across I\,Zw~18, revealing multiple clusters in which 
carbon-sequence Wolf-Rayet stars were observed. Such prominent WR 
signatures at low metallicity led Crowther \& Hadfield (2006)\nocite{ch06} 
to conclude 
that I\,Zw~18 hosts substantial WR populations, wholly unexpected for 
single 
star evolutionary models at such low metallicity.

The composite rest-frame UV spectra of z$\sim$3 Lyman break galaxies (LBG)
(Shapley et al. 2003)\nocite{sha03} also includes spectral signatures of O 
and WR stars.
Amongst the brightest (lensed) LBGs, MS1512-cB58 (z$\sim$2.7) has been
observed at rest-frame UV wavelengths with Keck, with which  Pettini et 
al. (2002)\nocite{pet02} analysed the interstellar lines, suggesting a 
Magellanic Cloud
metallicity. Indeed, Pettini et al. (2003)\nocite{pet03} could reproduce 
its 
stellar CIV 1550 feature using Magellanic Cloud template OB stars, where
they had previously been unsuccessful using Milky Way template stars.

\section{Future Issues?}

Perhaps the most significant unresolved aspect relating to high mass stars 
is their formation, for which future high-spatial resolution infrared and 
radio may prove decisive. Violent mass ejections, such as that experienced 
by $\eta$ Carinae, producing the Homunculus nebula, also remains 
ill-explained. A further puzzle is how large numbers of Wolf-Rayet stars 
form in young, massive clusters within very metal-poor galaxies such as 
I\,Zw~18.



\printindex
\end{document}